\documentclass{jltp}

\usepackage{graphicx} 
\usepackage{amssymb,amsmath,bm}
\newcommand{\subF}{\mathrm{F}}

\title{Overscreening Diamagnetism in Cylindrical Superconductor-Normal
  Metal-Heterostructures}
 
\author{W. Belzig, C. Bruder$^*$, and Yu.V. Nazarov$^\dagger$}

\address{Department of Physics, University of Konstanz, 
 78457 Konstanz, Germany\\
$^*$Department of Physics and Astronomy, University of Basel,\\ 
Klingelbergstrasse 82, 4056 Basel, Switzerland\\
$^\dagger$Kavli Institute of NanoScience, Delft University of
 Technology,\\ 2628 CJ Delft, The Netherlands}

\runninghead{W. Belzig, C. Bruder, and Yu.V. Nazarov}{Overscreening
  Diamagnetism in Cylinderical ...}

\begin{document}

\maketitle

\begin{abstract}
  We study the linear diamagnetic response of a superconducting
  cylinder coated by a normal-metal layer due to the proximity effect
  using the clean limit quasiclassical Eilenberger equations.  We
  compare the results for the susceptibility with those for a planar
  geometry. Interestingly, for $R\sim d$ the cylinder exhibits a
  stronger overscreening of the magnetic field, i.e., at the interface
  to the superconductor it can be less than (-1/2) of the applied
  field. Even for $R\gg d$, the diamagnetism can be increased as
  compared to the planar case, viz. the magnetic susceptibility
  $4\pi\chi$ becomes smaller than $-3/4$. This behaviour can be
  explained by an intriguing spatial oscillation of the magnetic field
  in the normal layer.
\end{abstract}

\section{Introduction}

A normal metal in electronic contact to a superconductor acquires
superconducting properties. This phenomenon is called proximity 
effect\cite{degennes:64,orsay:66}.  One of these properties is the
diamagnetic screening of an applied magnetic field, which has been
studied in a series of 
experimental\cite{mota:82,pobell:87,mota:89,mota:90,mota:94,stalzer:06} and
theoretical 
works\cite{zaikin:82,higashitani:95,onoe:95,belzig:96,fauchere:97,belzig:98}.

First predictions on the induced screening properties were made on the
basis of the Landau-Ginzburg theory by de Gennes and 
coworkers\cite{degennes:64,orsay:66}. The proximity effect is governed by the
thermal coherence length $\xi_T$, given by $\sqrt{D/T}$ in the
diffusive regime and $v_{\subF}/T$ in the clean limit. Here $v_{\subF}$ denotes
the Fermi velocity and $D$ the diffusion coefficient inside the normal
metal.  Throughout the paper, we will set $\hbar=k_{\mathrm{B}}=1$. 
The Ginzburg-Landau approach is valid outside the mesoscopic
regime in which $\xi_T\ll d$, where $d$ is the thickness of the normal
metal film. In this approach a region of width $\sim\xi_T$ of the normal metal
screens the magnetic field and consequently the susceptibility of the
normal metal $4\pi\chi=4\pi M/H=-\xi_T/d$ increases with the coherence
length. Strictly speaking the approach of de Gennes {\it et al.}
requires in addition a local constitutive relation
$j(x)\sim A(x)$\cite{london:38}, which is only the case under special
circumstances\cite{belzig:98}.
The first experimental studies were in agreement with these early
predictions\cite{oda:80,oda:84} and the predicted screening was
observed. However, in this experiment the mesoscopic regime was not
reached. According to the prediction the susceptibility approaches
almost the ideal value $4\pi\chi=-1$.

In 1980 Zaikin investigated the screening properties of clean normal
metals in the clean limit. In that case the constitutive relation is
complete non-local, viz. $j\sim\int_0^dA(x)dx$. This has the
interesting consequence that the susceptibility saturates at low
temperatures at $75\%$ of the ideal diamagnetic value. The magnetic
field inside the normal metal decays linearly at low temperatures and
even changes sign to reach a maximal opposite field of $-H/2$ at the
interface to the superconductor. Note, that a similar effect occurs in clean
type-I superconductors\cite{pippard:53}.

Around the same time a series of experiments by Mota and coworkers
found an interesting low-temperature anomaly in the magnetic response
of cylindrical structures
\cite{mota:82,mota:89,mota:90,mota:94,mueller:00,mueller:00a}. At very
low temperatures the diamagnetic signal decreased again and in the end
became even paramagnetic. This so-called reentrant effect of the
magnetic susceptibility has triggered a number of theoretical
explanations\cite{bruder:99,fauchere:99,belzig:00,niederer:02},
however a final experimental verification of one proposal is still
missing.

The effect of elastic impurity scattering was the subject of a number
of works. A finite elastic mean free path reduces the range of the
current-field relation and changes the screening properties
drastically\cite{belzig:98}. Remarkably, impurities can enhance the
screening ability at some temperatures and even have an effect if the
mean free path exceeds the thickness of the normal layer. This
quasiclassical description\cite{belzig:99} was shown to agree with the
experimental data for temperatures above the reentrance
regime\cite{mueller:99}. In a number of other works the nonlinear
magnetic properties have been discussed numerically \cite{belzig:96}
and analytically\cite{fauchere:97}. Other works addressed the effect
of a non-ideal interface
\cite{ashida:89,narikiyo:89,galaktionov:02,galaktionov:03}.

\begin{figure}
  \begin{center}
    \includegraphics[width=5cm]{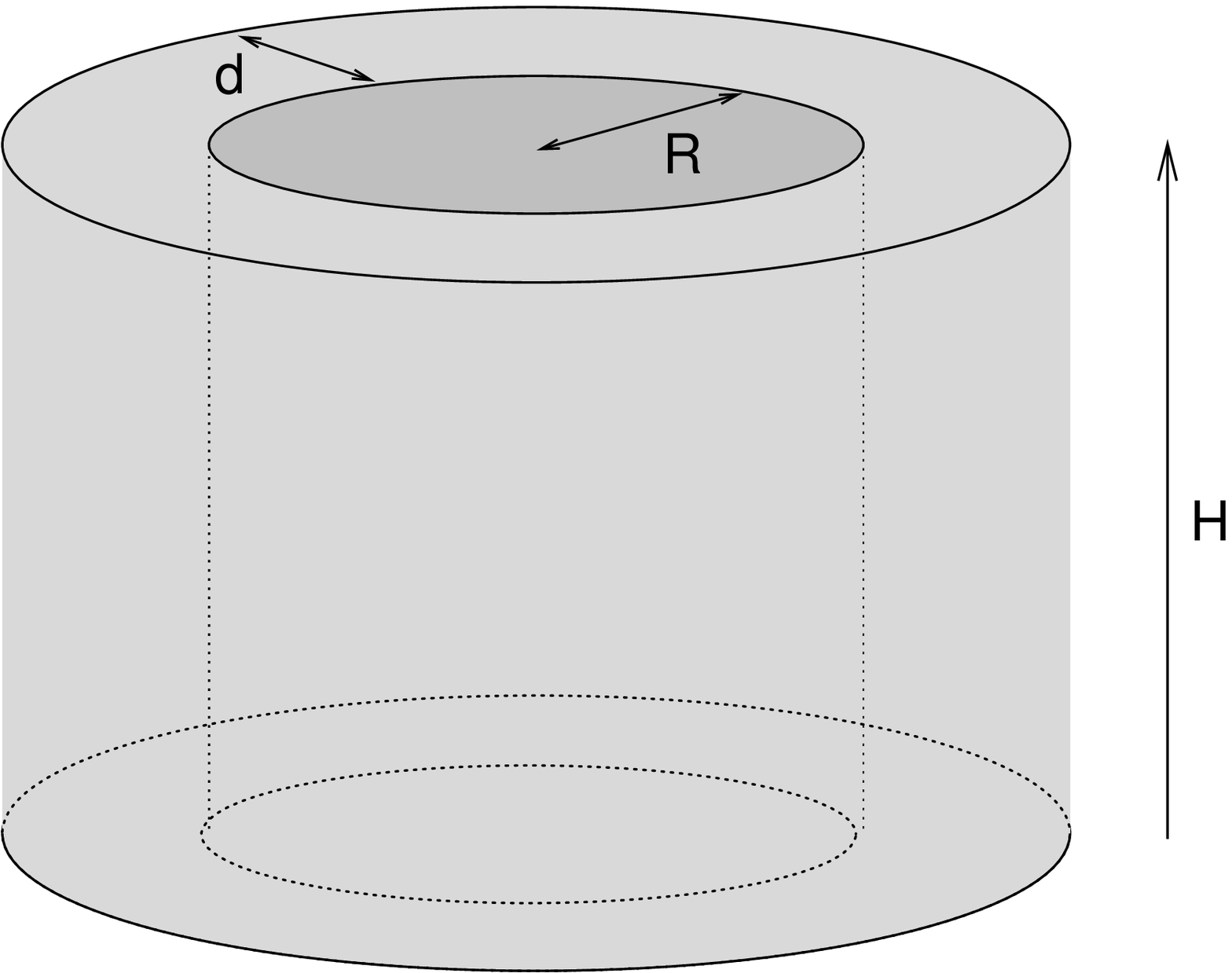}
    \includegraphics[width=6cm]{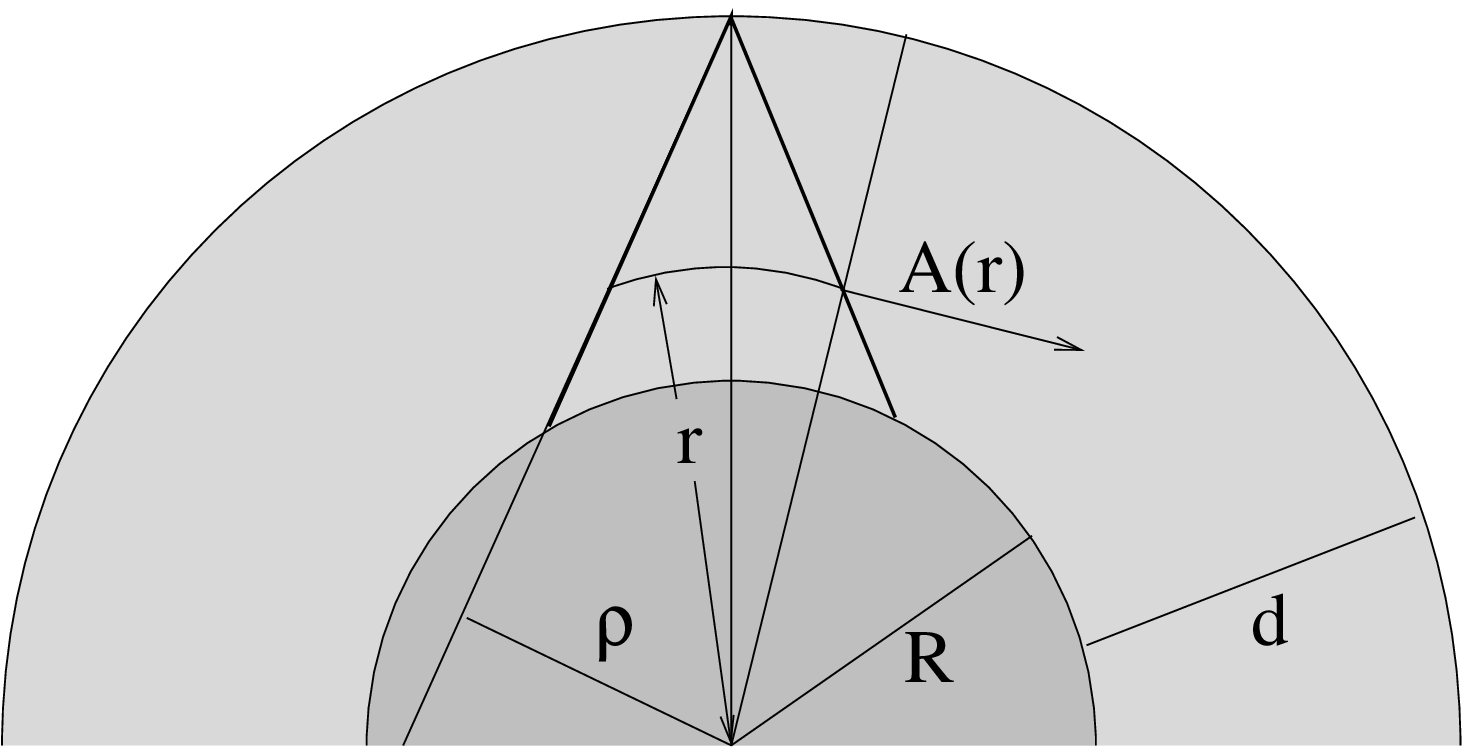}
  \end{center}
  \caption{Left panel: a superconducting cylinder surrounded by a
  normal-metal layer. A magnetic field is applied in parallel to the
  axis of the cylinder. Right panel: 
  parameterization of the averaging procedure over the Fermi surface.}
\label{param}
\end{figure}

In this work we investigate the effect of a cylindrical geometry on
the magnetic screening properties. A sketch of the system is shown in
the left panel of Fig.~\ref{param}. 
We will consider a superconducting cylinder covered by a normal-metal
layer.  The superconducting core has radius $R$; the normal-metal
layer has thickness $d$, i.e., the total radius of the cylindrical
sample is $R+d$. Both the superconductor and the normal metal are
assumed to be in the clean limit. A magnetic field is applied parallel
to the axis of the cylinder, and the task is to calculate the 
magnetic susceptibility of the normal-metal layer 
(we assume an ideal diamagnetism of the superconductor).

We assume the pair potential $\Delta$ to have a step-like dependence
on the radial coordinate,
\begin{equation}
\label{Paarpotential}
\Delta (r)=\Delta \Theta (R-r)\; ,
\end{equation}
and we will neglect self-consistency. 
This is justified since the superconductor is assumed to be much
thicker than the coherence length and the magnetic field is assumed to
be much less than the critical field. A small suppression of $\Delta$
at the interface would not lead to qualitative changes. 

Throughout this paper, we will use the symmetric gauge for the vector
potential
\begin{equation}
\label{Eichung}
\bm{A}(\bm{r})=A(r)\bm{e}_\varphi\; ,\quad
\bm{H}(\bm{r})=H \bm{e}_z \Theta(r-R-d)\; .
\end{equation}
Here $\bm{e}_\varphi$ is the unit vector in direction of $\varphi$ and
$\bm{e}_z$ is the unit vector in $z$-direction.

The rest of the paper is organized as follows: in Section
\ref{sec:quasi}, we introduce the formalism of quasiclassical Green's
functions and discuss how to solve the Eilenberger equation. Its
solution leads to an expression for the supercurrent density in the
normal layer, that is combined with Maxwell's equation in Section 
\ref{sec:maxwell} which results in an integro-differential equation
for the vector potential. We solve this equation numerically in
Section \ref{sec:disc} and calculate the magnetic susceptibility of
the normal layer. In the planar limit, the results agree with earlier
work, whereas in the cylindrical case we find new and interesting
screening behavior and a non-trivial oscillating field distribution.

\section{Solution of  the Quasiclassical Equations}
\label{sec:quasi}

The goal is to solve the clean-limit Eilenberger equation
\cite{eilenberger:68,larkin:68}  
\begin{equation}
\label{app.eilenberger}
-\bm{v}_\subF\bm{\partial}\hat{g}_{\omega}(r,\bm{v}_\subF)=
[\:(\omega+ie \bm{v}_\subF\bm{A}(r))\hat\tau_3+\hat{\Delta}(r)\:,
\:\hat{g}_{\omega}(r,\bm{v}_\subF)\:] 
\end{equation} 
in the cylindrical geometry described above. Here $\omega=\pi T
(2n+1)$ denotes a Matsubara frequency and
$\hat\Delta=\Delta\hat\tau_1$ is the off-diagonal pair potential of
the superconductor. We will assume the following boundary conditions:
a fully transparent boundary at the NS interface,
\begin{equation}
\label{app.erbc1} 
\hat{g}_\omega^s (R-0,\bm{v}_\subF)=\hat{g}_\omega^n
(R+0,\bm{v}_\subF) 
\end{equation}
and specular reflection at the outer boundary
\begin{equation}
\label{app.erbc2} 
\hat{g}_\omega^n (R+d,\bm{v}_\bot,\bm{v}_\|)=
\hat{g}_\omega^n (R+d,-\bm{v}_\bot,\bm{v}_\|) 
\; .
\end{equation}
Deep in the superconductor, we have 
\begin{equation}
\label{app.erbc3} 
\hat{g}\rightarrow \frac{1}{\Omega}
\left(\begin{array}{cc} \omega & \Delta\\ \Delta & -\omega
\end{array}\right)
\end{equation} 
where $\Omega=\sqrt{\Delta^2+\omega^2}$. These equations have to be
solved along classical trajectories, along which they become effectively
one-dimensional.
The general solution is a superposition of the homogeneous solution
$\hat{g}_h$ and the rising resp. decaying solutions $\hat{g}_+$ and
$\hat{g}_-$. In the normal layer, $ R\le
 r\le R+d$ we make the Ansatz
\begin{equation} 
\hat{g}(r)^{n}=\alpha_0\,\hat{g}_h^{n}+
\alpha_+\,\hat{g}_+^{n}\,e\,\raisebox{2ex}{$\eta(r-R)$}+
\alpha_-\,\hat{g}_-^{n}\,e\raisebox{2ex}{$-\eta(r-R)$}\; , 
\end{equation}
where
\begin{equation}
\eta(x)=\frac{2\omega}{|v_{\subF x}|}\,x+2ie\frac{v_{\subF y}}{|v_{\subF x}|}
\int_0^{x}A(R+x^\prime)\,dx^\prime
\end{equation}
is the complex phase accumulated along the trajectory and
\begin{equation}
\hat{g}_h^n = \tau_3\; ,\quad
\hat{g}_{\pm}^n = \frac{1}{2}(\tau_1\mp i \,\mathrm{sign}
(v_{\subF x})\tau_2)\,.
\end{equation}
In the superconductor, the corresponding Ansatz reads
\begin{equation}
\label{Allgloesungs}
\hat{g}(r)^{s}=\beta_0\,\hat{g}_h^{s}+
\beta_+\,\hat{g}_+^{s}\,e\,\raisebox{2ex}{$\frac{2\Omega
    (r-R)}{|v_{\subF x}|}$}+
\beta_-\,\hat{g}_-^{s}\,e\,\raisebox{2ex}{$-\frac{2\Omega
    (r-R)}{|v_{\subF x}|}$}\; , 
\end{equation} 
where
\begin{equation}
\hat{g}^s_h = \frac{1}{\Omega}(\omega\tau_3+\Delta\tau_1)\; ,\quad
\hat{g}_{\pm}^s = \frac{1}{2\Omega}(\Delta\tau_3-\omega\tau_1
        \pm i\Omega\,\mathrm{sign} (v_{\subF x})\tau_2)\;.
\end{equation}
We have neglected the magnetic field in the superconductor, since the
penetration depth is assumed to be much smaller than the coherence
length. Furthermore, we will neglect the possibility that a
quasiparticle trajectory leaves the superconducting core again.

The coefficients $\alpha_0,\beta_0,\alpha_{+/-}$ and $\beta_{+/-}$
fulfill the normalization conditions
\begin{equation}
\alpha_0^2+\alpha_+\alpha_-=1\quad,\quad
\beta_0^2+\beta_+\beta_-=1\; . 
\end{equation}
Using the boundary conditions, we find that
\begin{equation}
\beta_-=0 \quad \textrm{and}\quad
\beta_0=1
\end{equation}
deep in the superconductor, whereas the boundary condition at $x=R+d$
in the normal conductor leads to
\begin{equation}
\alpha_+\,e\raisebox{2ex}{$\eta(d)$}= \alpha_-\,e\raisebox{2ex}{$-\eta(d)$}
\; .
\end{equation}
The other two coefficients can be determined from the matching
condition at $r=R$  leading to the final result
\begin{equation} 
\alpha_0 = \frac{\Omega+\omega\coth(\eta(d))}{\omega+\Omega\coth(\eta(d))}
\; ,\quad 
\beta_+ = \frac{2\Delta}{\omega+\Omega\coth(\eta(d))}\; .
\end{equation}
The diagonal component of the Green's function in the normal metal
($\tau_3$-component of $\hat{g}^n$) is constant along a trajectory,
\begin{equation}
        g({\cal L})=\frac{\Omega+{\omega}\coth(\eta(d))}
        {{\omega}+\Omega\coth(\eta(d))},
\end{equation}
where
\begin{equation}
        \eta=\frac{{\omega} L_{\cal L}}{v_\subF}
        +ie\int_{\cal L}\bm{A}(\bm{r})\bm{dr}
\end{equation}
and
\begin{equation}
        \Omega=\sqrt{\Delta^2+\omega^2}\; .
\end{equation}
Here, $L_{\cal L}$ is the length of trajectory ${\cal L}$ and
$\int_{\cal L}$ the line integral along the trajectory. Here we have
generalized our previous formula to the more general cylindrical
case. In the planar limit $R/d \to \infty$, our solution agrees with
earlier work\cite{zaikin:82}.

This solution determines the screening current in the normal metal
layer via
\begin{equation}
  \label{eq:current}
  \bm j(\bm r) = -i 4\pi e N_0 T\sum_{\omega>0}\langle \bm v_\subF
  \mathrm{Tr}\hat\tau_3\hat g_\omega(\bm v_\subF)\rangle=
  -i8\pi eN_0T\sum_{\omega>0}\langle\bm v_{\subF}\alpha_0\rangle\; .
\end{equation}
Here, $N_0$ denotes the density of states at the Fermi energy. We
note, that the screening current is determined by the diagonal
component of $\hat g$ only, which for a given trajectory is constant
along that particular trajectory. However, the angular average may
induce a space dependence, since certain trajectories, which do not
hit the superconductor have to be excluded from the angular
average. This will be important for the cylindrical geometry
considered below.

\section{Solution of Maxwell's equation}
\label{sec:maxwell}

We will solve Maxwell's equation to obtain the spatial dependence of
the magnetic field in the normal metal from the screening-current
density. In the planar limit $R/d \to \infty$, it is possible to give an
analytical solution, which we discuss briefly\cite{zaikin:82}. In this
case, the current density in the normal layer turns out to be constant
as a function of the spatial coordinates,
\begin{eqnarray}
\label{rein.strom.exakt}
        j & \stackrel{R/d\to\infty}{=} &
        16eN_0v_\subF\,T\sum_{\mu=0}^\infty 
        \int_0^{\frac{\pi}{2}}d\theta
        \int_0^{\frac{\pi}{2}}d\varphi\sin^2\theta\cos\varphi\\
\nonumber&& 
        \times \frac{\Delta^2(\cosh\Phi+\cos\phi)\,\sin\phi}
        {\left[{\omega}\sinh\Phi+\Omega\,
        (\,\cosh\Phi+\cos\phi\,)\right]^2}\; .
\end{eqnarray}
Here, 
\begin{equation}
        \Phi=\frac{2\omega L}{v_\subF}
        =\frac{4\omega d}{v_\subF\cos\theta}
\end{equation}
is an effective trajectory length, and 
\begin{equation}
        \phi=4\tan\theta\cos\varphi \,e\int_R^{R+d}A(r)dr
\end{equation}
is the Aharonov-Bohm phase along this trajectory.  The current density
depends on temperature and applied magnetic field $H$. More precisely,
there is a non-local relationship between current density and vector
potential that can be understood in the following way: consider the
semiclassical trajectory of an electron in the normal metal starting
at the NS interface with velocity $\bm{v}_\subF$.  It will be
specularly reflected at the outer interface and eventually hit the
superconductor, see the right panel of Fig.\ref{param}.  There it will
be Andreev-reflected and return as a hole along the trajectory that
the electron took before.  Such semiclassical closed orbits correspond
to quantum bound states and are known as Andreev
levels\cite{andreev:66}. In the presence of a vector potential, the
electron picks up a phase
\begin{equation}
e\int_L \bm{A}(\bm{r})\bm{dr}=
2e\frac{v_{\subF\varphi}}{v_{\subF r}}\int_R^{R+d}A(r)dr\; ,
\end{equation}
which will lead to an energy shift for the Andreev levels\cite{zaikin:81}
\begin{equation}
\frac{ev_{\subF\varphi}}{d}\int_R^{R+d}A(r)dr\; .
\end{equation}
Since the Andreev levels are homogeneous in the normal layer, 
this energy shift leads to a spatially constant current density.

Solving Maxwell's equation with the boundary conditions
\begin{equation}
        A(R)=0 
\label{bc1}
\end{equation}
and
\begin{equation}
\left.\frac{d}{dr}A(r)\right|_{r=R+d}+\frac{A(R+d)}{R+d}=H
\label{bc2}
\end{equation}
leads to the following $z$-component of the magnetic field:
\begin{eqnarray}
\label{feldrein}
B(r)=H-4\pi j(T,H) (R+d-r)&\mathrm{for}& R<r<R+d\\
\nonumber
B(r)=H&\mathrm{for}&r>R+d\; .
\end{eqnarray}
The field may change its sign within the normal-metal layer if 
\begin{equation}
\label{rein.ungleich}
4\pi j(T,H) d>H\; .
\end{equation}
This ``overscreening'' effect was first discussed by
Zaikin\cite{zaikin:82}.  A numerical evaluation of the current density 
Eq.~(\ref{rein.strom.exakt}) at all temperatures
provides a solution of Maxwell's equations for arbitrary fields.

For a cylindrical geometry, the current density in the normal layer
depends on the radial coordinate. 
The $\varphi$-component of the current density reads 
\begin{equation}
\label{app.current1}
j(r)=-ieN_0T\sum_{\omega>0}\langle 
v_{\subF\varphi}\alpha_0\rangle(r)\; .
\end{equation}
The average over the Fermi surface can be parameterized
in a way that allows to easily exclude trajectories that do not meet
the superconductor, i.e., do not contribute to the current.
Instead of performing the usual angular average
\begin{equation}
\int\frac{d\Omega_{v_\subF}}{4\pi}\cdots=\int_{-1}^1\frac{d\cos{\theta}}{2}
\int_0^{2\pi}\frac{d\varphi}{2\pi}\cdots
\end{equation}
it is convenient to define $\rho=r\sin{\varphi}$ which leads to 
\begin{equation}
\int_{-r}^{r}\frac{d\rho\,\rho}{r\sqrt{r^2-\rho^2}}
\int_{-1}^1\frac{d\cos{\theta}}{4\pi}\cdots\; ,
\end{equation}
see the right panel of Fig.~\ref{param} for the geometrical
interpretation of $\rho$.  Obviously, only those trajectories
contribute for which $|\rho|<R$.  Such a trajectory has length
\begin{equation}
L(\rho,\theta)=
2\frac{\sqrt{(R+d)^2-\rho^2}-\sqrt{R^2-\rho^2}}{|\sin{\theta}|}\; .
\end{equation}
Using this parameterization, the line integral over the vector potential 
can be written as
\begin{equation}
\int\bm{A}(r)\bm{dx}= 2\int_{R}^{R+d}dr^\prime A(r^\prime)
\frac{\rho}{\sqrt{r^{\prime 2}-\rho^2}}\,\mathrm{sign}\theta\; .
\end{equation}
We finally obtain the following expression for the current 
density, Eq.~(\ref{app.current1}), 
\begin{eqnarray}
j(r)&=&4eN_0v_\subF\;T\sum_{\omega>0}\,\int_0^Rd\rho\int_0^{\pi/2}d\theta
\frac{\rho\,\sin^2\theta}{2\pi r\sqrt{r^2-\rho^2}}\\
\nonumber&&
\hfill\times\,\frac{\Delta^2\sin\phi\,(\,\cosh\Phi+\cos\phi\,)}
{\left(\,\omega\sinh\Phi+\Omega\,
(\,\cosh\Phi+\cos\phi\,)\,\right)^2+\omega^2\sin^2\phi}\;,
\end{eqnarray}
where
\begin{equation}
\Phi=\frac{2\omega L(\rho,\theta)}{v_\subF}\;,
\end{equation}
and
\begin{equation}
\phi=4e\int_{R}^{R+d}dr^\prime A(r^\prime)
\frac{\rho}{\sqrt{r^{\prime 2}-\rho^2}}\; .
\end{equation}
 
Using Maxwell's equations, and to linear order in the magnetic field,
we find the following linear integro-differential equation for the
vector potential:
\begin{eqnarray}
\label{rein.maxwell}
\lefteqn{\xi^2\,\frac{d}{dr}\,\frac{1}{r}\,\frac{d}{dr}\,r\,A(r) =
        K\,\frac{T}{T_c}\,\sum_{{\omega}>0}\,
        \int_0^R\frac{d\rho}{\xi}\int_0^{\pi/2}d\theta}\\
& & \nonumber
        \times\,\frac{\rho^2\,\sin^2\theta}{r\sqrt{r^2-\rho^2}}\,
        \frac{\Delta^2\,(\,\cosh\Phi+1\,)}
        {\left(\,{\omega}\sinh\Phi+\Omega\,(\,\cosh\Phi+1\,)\,\right)^2}
        \int_R^{R+d}\frac{A(r^\prime)\,dr^\prime}{\sqrt{r^{\prime
        2}-\rho^2}}\; .
\end{eqnarray}
Here, the effective length of the trajectories is given by 
\begin{equation}
        \Phi=
        \frac{4\omega(\sqrt{(R+d)^2-\rho^2}-\sqrt{R^2-\rho^2})}
        {v_\subF|\sin{\theta}|}\; .
\end{equation}
We have introduced dimensionless units; lengths are measured in units of
the (clean-limit) coherence length at $T_c$
\begin{equation} 
        \xi =\frac{v_\subF}{2\pi T_c}\; ,
\end{equation}
the current density contains a dimensionless material constant,
\begin{equation}
K=32\,e^2N_0v^2_\subF\xi^2=
\frac{24}{\pi} \left(\frac{\xi}{\lambda_N}\right)^2\; ,
\label{material_parameter}
\end{equation}
where $\lambda_N=(4\pi e^2n_e/m)^{-1/2}$ is a length scale that is
defined in analogy with the London penetration depth, but using the
electron density $n_e$ of the normal metal instead of the superfluid
density of the superconductor. 

\section{Results for the magnetic susceptibility}
\label{sec:disc}

We will now discuss the susceptibility in the linear-response regime
for the planar and cylindrical case.

\begin{figure}
  \begin{center}
\includegraphics[width=12cm]{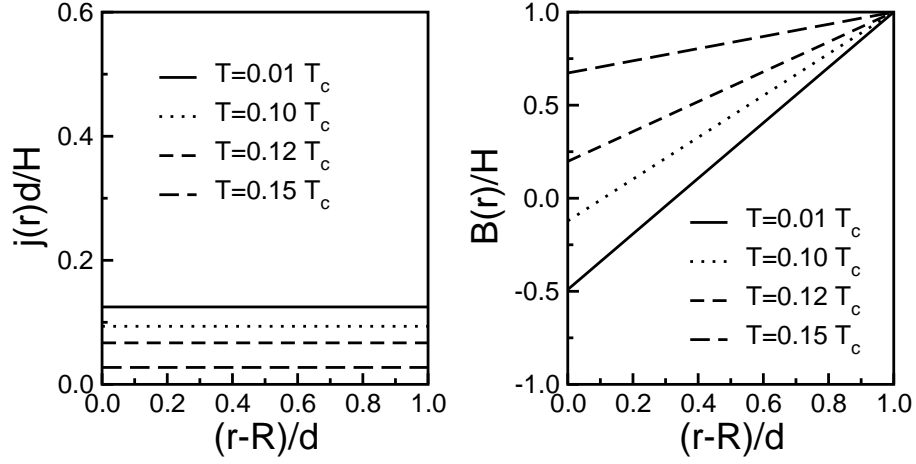}
  \end{center}
\caption{\label{r20000} Current density (left panel) and magnetic
field (right panel) for a planar geometry with $\xi=d/20$,
$R/d \to \infty$, $K=10$, and different temperatures. The material parameter
$K$ is defined in Eq.~(\ref{material_parameter}).}
\end{figure}

The numerical solution in the planar case for $K=10$ and a normal-layer
thickness of $20\,\xi$ leads to the current densities and
magnetic fields shown in Fig.~\ref{r20000}. The left panel shows the
current density (that is constant for the planar case). The right
panel shows the corresponding spatially dependent magnetic fields.
For $T\lesssim 0.1 T_c$ the condition (\ref{rein.ungleich})
is fulfilled, and the field changes sign at the NS interface. Below
$T=0.05T_c$ the current density does not increase any more, and we
only show the curve for $T=0.01T_c$.

\begin{figure}
  \begin{center}
\includegraphics[width=12cm]{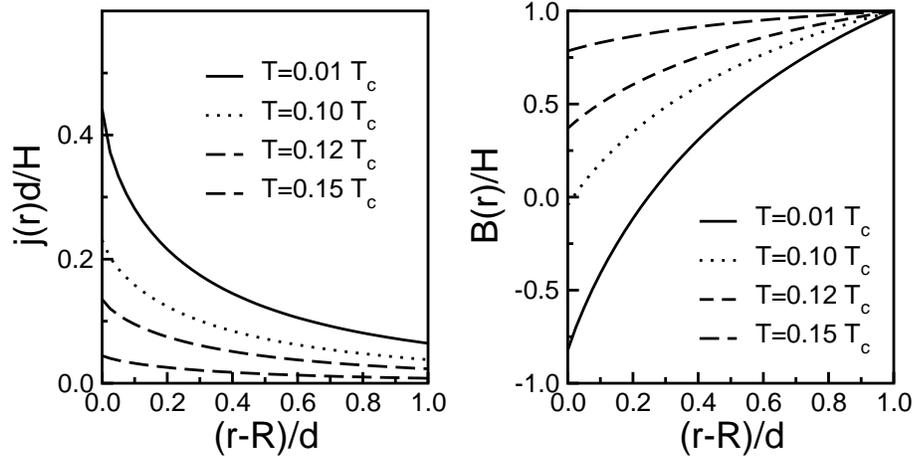}
  \end{center}
\caption{\label{r20}
Current density (left panel) and magnetic field (right panel) for a
cylindrical geometry with $d=R$, $\xi=d/20$, $K=10$, and 
different temperatures.}
\end{figure}

As discussed before, the current density depends on the spatial
coordinate in the cylindrical case. Figure~\ref{r20} shows the current
density and the magnetic field for 
$K=10$, both the radius $R$ of the superconductor and the normal-layer
thickness have been chosen as $20\,\xi$.  
The local current density can be significantly larger than in the
planar case, e.g., three times as large at the NS interface for 
$T\rightarrow 0$.  
The screening effect is weaker than in the planar case close to the
outer interface; close to the NS-interface it is increased, and the
overscreening effect may be even larger than in the planar case.

The susceptibility of the normal-metal layer can be obtained from the
solution of Maxwell's equation via 
\begin{equation}
\label{Suszrein}
4\pi\chi(T,H)=\frac{A(R+d)}{A_e(R+d)}-1\; ,
\end{equation}
where $A_e(r)$ 
is the vector potential {\it in the absence} of the normal layer.
In the planar case, the vector potential can be calculated
analytically and leads to
\begin{equation}
4\pi\chi(T,H)=-2\pi\frac{j(T,H)d}{H}\; .
\end{equation}
We will now give explicit expressions for the susceptibility in the
linear-response regime for the planar and cylindrical case.

In the planar case, Eq.~(\ref{rein.strom.exakt}) 
can be simplified assuming $e\int_R^{R+d}A(r)dr\ll 1$ or 
$H\ll \Phi_0/d^2$. In other words, the flux enclosed by a
semiclassical trajectory is less than a flux quantum $\Phi_0$; the
phase picked up by the electron has to be less than $2\pi$. 
For experimentally relevant normal-layer thicknesses of 
$d\approx 1\mu$m these conditions are fulfilled for 
fields $H\le 20$G.  
The current density (\ref{rein.strom.exakt}) factorizes like
\begin{equation}
\label{rein.schwachstrom}
        j(T,H)=-ej_s(T)\int_R^{R+d}A(r)dr\; ,
\end{equation}
where
\begin{eqnarray}
\lefteqn{j_s(T)=4\pi e N_0v_\subF T\,\sum_{\mu=0}^\infty
        \int_0^{\frac{\pi}{2}}d\theta\sin^2\theta\tan\theta}\\
        \nonumber & & \qquad\qquad
        \frac{\Delta^2\left
        (\cosh\left(\frac{4\omega d}{v_\subF\cos\theta}\right)+1\right)}
        {\left[\omega\,
        \sinh\left(\frac{4\omega d}{v_\subF\cos\theta}\right)+
        \Omega\,
        \left(\cosh\left(\frac{4\omega
        d}{v_\subF\cos\theta}\right)+1\right)
        \,\right]^2}\; .
\end{eqnarray}
The analytical solution of Maxwell's equation, Eq.~(\ref{feldrein}) 
leads to the following expression for the susceptibility
\begin{equation}
\label{schwachsusz}
4\pi\chi(T)=\frac{-3\pi ej_s(T)\,d^3}{3+4\pi ej_s(T)d^3}\; .
\end{equation}
In general, $j_s$ can be calculated only numerically.
However, for $T=0$ an approximate analytical expression can be
obtained, 
\begin{equation}
\label{T=0}
j_s(T=0)\approx \frac{2}{\pi^3ed\lambda_N^2}\; .
\end{equation}
For $d\gg \lambda_N$, Eq.~(\ref{schwachsusz}) leads to a
susceptibility that it is independent of geometry,
\begin{equation}
\label{chi0}
4\pi\chi(T=0)=-\frac{3}{4}\; .
\end{equation}
Field expulsion is never complete; the maximal value that the
susceptibility can reach at any temperature is $3/4$ of an ideal
diamagnet.\cite{zaikin:82}

\begin{figure}
  \begin{center}
\includegraphics[width=6cm]{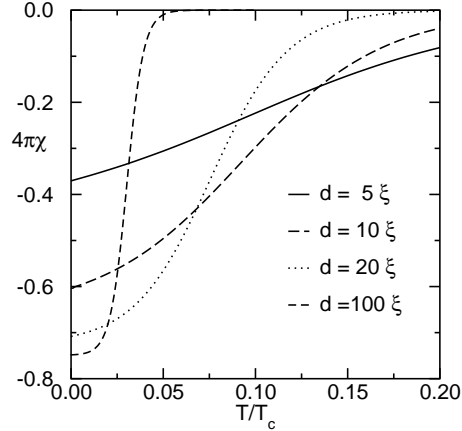}
  \end{center}
\caption{\label{rein.versch.d}
Susceptibility of the normal-metal layer as a function of 
its thickness $d$ for $R/d\to \infty$ and $K=1$.}
\end{figure}

Figure~\ref{rein.versch.d} shows the susceptibilities for different
normal-layer thicknesses. In thicker layers, field expulsion occurs
only at lower temperatures. The thickest layer reaches the saturation
value given in Eq.~(\ref{chi0}) for $T\rightarrow 0$, 
since it fulfills $d\gg\lambda_N$.  

For the cylindrical case, the integro-differential equation for the
vector potential, Eq.~(\ref{rein.maxwell}), has to be solved
numerically.

Surprisingly, at low temperatures and for
$K \gg 1$, the susceptibility does {\it not} go to $-3/4$. To analyze
this parameter range in a quantitative way, we start from
Eq.~(\ref{rein.maxwell}).  We assume $d \ll R$ and concentrate on the
contribution of the (long) trajectories with $R-\rho \simeq d$. The
typical length of these trajectories is $L \simeq \sqrt{d R}$. They
become important at low temperatures $T \ll v_\subF/\sqrt{dR}$. 
We will assume the zero-temperature limit and replace the sum over
$\omega$ in Eq.~(\ref{rein.maxwell}) by an integral.
This integral converges at $\omega \simeq v_\subF/L \ll \Delta$, and the
integration is readily performed.  The next step is to implement the
condition $d\ll R$.  To this end, we do the following substitutions:
\begin{equation}
\rho = R -\mu d \;,
\end{equation}
and
\begin{equation}
r = R +z d\;; \quad r'=R+z'd\;,
\end{equation}
where $\mu$, $z$, $z'$ are dimensionless variables with $\mu >0$ and
$0<z,z'<1$.  For instance, this gives $\sqrt{r'2 -\rho^2} \to
\sqrt{2dR} \sqrt{z'+\mu}$.  
Here and in the following, we will write $A(z)$ instead of $A(R+zd)$.
After this substitution,
Eq.~(\ref{rein.maxwell}) takes the following form:
\begin{equation}
\left(\frac{\xi}{d}\right)^2 \frac{d^2 A} {dz^2} 
= \frac{K}{12\sqrt{2}} \left( \frac{d}{R}\right)^{1/2}
\int_0^{\infty} d\mu \int_0^{1} dz'
\frac{A(z')}{(\sqrt{\mu+1}-\sqrt{\mu}) \sqrt{z+\mu} \sqrt{z'+\mu}}\;.
\label{inf}
\end{equation}
There is a subtle point here: 
the integral over $\mu$ diverges at big $\mu$.
Therefore, the right-hand side (r.h.s.) cannot describe the constant part of
the current density correctly, however, it gives the
correct inhomogeneous part. To get a physical result we perform the
following renormalization procedure: we introduce
a yet unknown constant $C$ on the r.h.s. to reproduce the constant part
of the density and rewrite Eq.~(\ref{inf}) as
\begin{multline}
\left(\frac{\xi}{d}\right)^2 \frac{d^2 A(z)} {dz^2} = C + 
\frac{K}{12\sqrt{2}} \left(\frac{d}{R}\right)^{1/2}\int_0^{\infty} 
d\mu \int_0^{1} dz'\\A(z')
\left(\frac{1}{(\sqrt{\mu+1}-\sqrt{\mu}) \sqrt{z+\mu} \sqrt{z'+\mu}} -
4(\sqrt{\mu+1}-\sqrt{\mu})\right)\;.
\label{infmu}
\end{multline}
The $\mu$-dependent expression has been subtracted
from the second term on the r.h.s. to assure that after
subtraction the term gives zero if integrated over $z$, $z'$.
The integral over $\mu$ converges now. (We could use any other
form of this subtraction that provides convergence of the integral
since the change would be incorporated in $C$).

We do not have to know the coefficient that
relates $C$ and $A$ provided that this coefficient is sufficiently
big, which is true for $\lambda \ll d$. 
In this limit, $C$ has to be determined from the extra condition
$\int dz A(z) = 0$. 

The remaining task is to numerically solve Eq.~(\ref{infmu}) under the
conditions
\begin{eqnarray}
\int_0^1 dz A(z) =0\;; \\
\frac{d}{dz} A(z)|_{z=1} = Hd\;;\\
A(0) =0\;.
\label{infmubc}
\end{eqnarray}
The susceptibility is determined from this solution
as
\begin{equation}
4\pi \chi = A(1)/Bd-1\; .
\label{infmuchi}
\end{equation}
Looking at Eq.~(\ref{infmu}) we see that 
the susceptibility is a function of the parameter
\begin{equation}
\gamma \equiv 1/\sqrt{ K (d/\xi)^2 (d/R)^{1/2}} 
=\sqrt{\frac{\pi}{24}} (\lambda/d)(R/d)^{1/4}\;.
\label{gamma}
\end{equation}
If $\gamma \gg 1$, $4\pi\chi$
goes to the known value of $-3/4$ and the spatial profile of $A$ is
determined by the magnetic energy. If $\gamma \ll 1$, the inhomogeneous
terms dominate and $4\pi\chi$ goes to $-1$. The crossover takes place at 
$\gamma \simeq 1$.

\begin{figure}
  \begin{center}
    \includegraphics[width=8cm,clip=true]{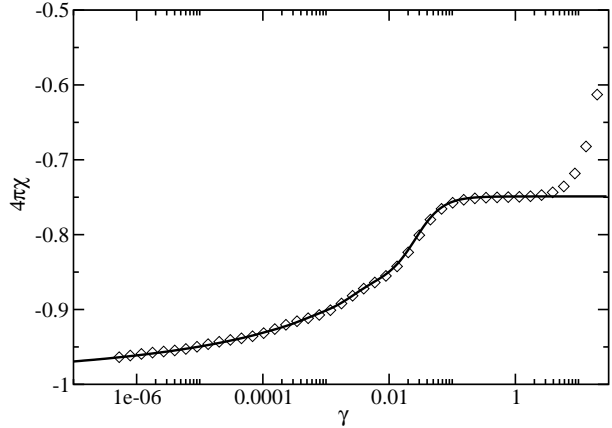}
  \end{center}
\caption{\label{figgamma} Susceptibility
  as a function of the parameter $\gamma$ defined in
  Eq.~(\ref{gamma}).
  Solid line: universal behavior predicted by 
Eqs.~(\ref{infmu}) - (\ref{infmuchi}). 
  Symbols: solution of Eqs.~(\ref{rein.maxwell}), (\ref{bc1}),
  (\ref{bc2}), and (\ref{Suszrein}) for $R/d=10^8$.}
\end{figure}

\begin{figure}
  \begin{center}
    \includegraphics[width=8cm,clip]{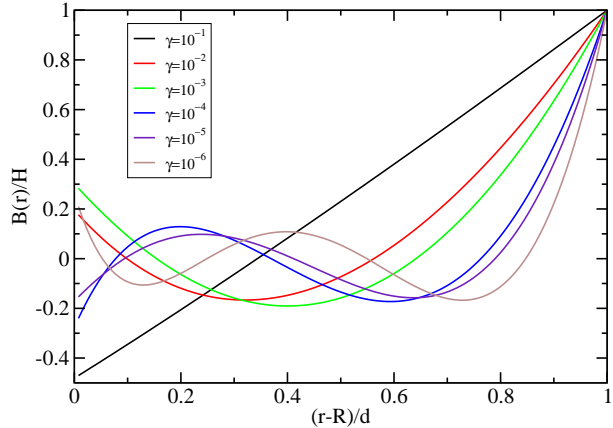}
  \end{center}
\caption{\label{fielddist} Magnetic field distribution in the normal
  layer for different values of $\gamma$.}
\end{figure}

Figure \ref{figgamma} shows the susceptibility as a function of the
parameter $\gamma$ defined in Eq.~(\ref{gamma}). The universal
behavior predicted by Eq.~(\ref{infmu}) and following (solid line) is
compared to a full numerical solution of the integro-differential
equation Eq.~(\ref{rein.maxwell}) for $R/d=10^8$ (symbols). 
The two curves agree perfectly for
$\gamma \lesssim 5$; the deviation for larger values of $\gamma$ is
explained by the fact that the condition $\lambda \ll d$ assumed in
the derivation of the universal curve is not fulfilled.  The
susceptibility shows a very interesting behavior. For small values of
$\gamma$ the susceptibility deviates substantially from the planar
case. In particular, its absolute value {\it increases}, i.e., the
screening is enhanced by the cylindrical geometry. This is in contrast
to the previously discussed behavior for $R\approx d$. The
difference can be understood by looking at the magnetic field inside the
normal metal, see Fig.~\ref{fielddist}. The spatially dependent
integral kernel allows the existence of higher Fourier components of the
current, instead of only the lowest-order component $j=$const. as in
the planar case. This leads to characteristic spatial oscillations 
of the magnetic field, which can show several sign changes inside the
normal-metal layer. In particular, the field at the interior interface
to the superconductor is also oscillating as a function of the
parameter $\gamma$. It is interesting to note that a similar phenomenon
occurs in the so-called Pippard superconductors\cite{pippard:53},
and has been experimentally observed\cite{drangeid:62}.

Finally, we would like to point out the qualitative similarity of the
effect discussed and the reduced diamagnetic response found in
Ref.~\onlinecite{belzig:98} for a planar geometry and anomalously
small impurity concentrations. In both cases, the origin of the effect
is the absence of the contribution of very long trajectories. While in
the cylindrical geometry the typical length scale of these
trajectories is $\sqrt{dR}$ ($\gg d$ for $d \ll R$), the presence of
impurities limits the trajectory length to the elastic mean free path.

\section{Conclusion}

We have investigated the induced magnetic screening properties of
cylindrical normal metal-superconductor heterostructures. The
ballistic screening leads to a peculiar non-local current-vector
potential relation inside the normal metal, governed by the parameter
$\gamma$ defined in Eq.~(\ref{gamma}). For $\gamma>1$ the screening is
like in the planar case, viz. the current density is constant, the
magnetic field decays linearly down to $-H/2$ and the susceptibility
is $4\pi\chi=-3/4$. In contrast, for $\gamma<1$, the field has an
oscillatory spatial dependence, including several sign changes inside
the normal metal. Overall, the magnetic susceptibility is enhanced and
reaches the ideal diamagnetic value $-1$ in the limit $\gamma\ll 1$. 

This work was financially supported the SFB 513 Nanostructures at
Surfaces and Interfaces of the DFG, the Swiss NSF, and the NCCR
Nanoscience.

\end{document}